\documentstyle[prd,aps,
multicol]{revtex}

\mathchardef\ddash="705C

\begin{document}

\title{Fate of Chiral Symmetries in Supersymmetric Quantum Chromodynamics
\footnote{Talk given by M.Y. at XXXth International Conference on High Energy 
Physics, July 27-August 2, 2000, Osaka, Japan}
}

\author{Yasuharu Honda$^a$ 
\footnote{E-mail: 8jspd005@keyaki.cc.u-tokai.ac.jp}
and Masaki Yasu${\grave {\rm e}}$$^{a,b}$
\footnote{E-mail: yasue@keyaki.cc.u-tokai.ac.jp}
}

\address{ $^{a}$ Department of Physics, Tokai University, Hiratsuka, Kanagawa 259-1292, JAPAN}
\address{ $^{b}$ Department of Natural Science School of Marine
Science and Technology,\\Tokai University, Shimizu, Shizuoka 424-8610, JAPAN}
\date{TOKAI-HEP/TH-0003, August, 2000}
\maketitle
\begin{abstract}
In supersymmetric quantum chromodynamics with $N_c$-colors and $N_f$-flavors
of quarks, our effective superpotential provides the alternative description to 
the Seiberg's $N=1$ duality at least for $N_f$ $\geq$ $N_c$+2, where spontaneous 
breakdown of chiral symmetries leads to $SU(N_c)_{L+R}$ $\times$ 
$SU(N_f-N_c)_L$ $\times$ $SU(N_f-N_c)_R$ as a nonabelian chiral symmetry.
The anomaly-matching is ensured by the presence of Nambu-Goldstone superfields 
associated with this breaking and the instanton contributions are properly equipped in 
the effective superpotential. 
\end{abstract}
%
\begin{multicols}{2}

\section{Prologue}

In supersymmetric quantum chromodynamics (SQCD) with $N_c$-colors and $N_f$-flavors 
of quarks, we have chiral $SU(N_f)$ symmetry.  At low energies, we have its dynamical breakdown to 
vectorial $SU(N_f)$ symmetry, for $N_f$ $\leq$ $N_c$. For remaining cases, we have 
restoration of chiral $SU(N_f)$ symmetry including the case with $N_f$ = $N_c$. 
For $N_f$ $\geq$ $N_c$+2, we need $\ddash$magnetic" degrees of freedom, namely, $\ddash$magnetic" 
quarks\cite{Seiberg}.  And, especially, for $3N_c$/2 $<$ $N_f$ $<$ 3$N_c$, the well-defined $N$ = 2 duality 
supports this description based on the $N$ = 1 duality\cite{CoulombPhase,BrokenN_2}.

In this talk, I will add the $\ddash$electric" description expressed in terms of mesons and baryons instead of 
$\ddash$magnetic" quarks to SQCD with $N_c$+2 $\leq$ $N_f$ $\leq$ 3$N_c$/2\cite{Yasue}.  In the $\ddash$electric" 
phase, however, since anomaly-matching\cite{tHooft} is not satisfied, we 
expect spontaneous breakdown of chiral symmetries\cite{AnomalyMatch}.  The residual symmetries will include vectorial 
$SU(N_c)$ symmetry and chiral $SU(N_f-N_c)$ symmetry, which are found by inspecting vacuum 
structure of our effective superpotential to be discussed.
 
\section{Anomalous $U(1)$ Symmetry and Superpotential}

We follow the classic procedure to construct our effective superpotential, which explicitly uses 
$S$ composed of two chiral gauge superfields\cite{VY}.  We impose on the superpotential the relation: 
$\delta{\cal L}$ $\sim$ $F^{\mu\nu}{\tilde F}_{\mu\nu}$ under the anomalous $U(1)$ transformation, 
where ${\cal L}$ represents the lagrangian of SQCD and $F^{\mu\nu}$  (${\tilde F}_{\mu\nu}\sim 
\epsilon_{\mu\nu\rho\sigma}F^{\rho\sigma}$) is a gluon's field strength. As a result, the anomalous 
$U(1)$-term is reproduced by the $F$-term of $S$.  

We find a superpotential, where mesons and baryons are denoted by $T$, $B$ and ${\bar B}$\cite{Weff}:
\begin{eqnarray}\label{Eq:Weff}
W_{\rm eff} &=& S 
\left\{ 
\ln\left[
\frac{
	S^{N_c-N_f}{\rm det}\left(T\right) f(Z)
}
{
	\Lambda^{3N_c-N_f}
} 
\right] \right.
\nonumber \\
&&+\left.N_f-N_c\right\}
\end{eqnarray}
with an arbitrary function, $f(Z)$, to be determined, where $\Lambda$ is the scale of SQCD and 
$Z$ is defined by (with abbreviated notations) $BT^{N_f-N_c}\bar{B}/{\rm det}(T)$.  
This is the superpotential to be examined. It looks familiar to you except for the function $f(Z)$ here. 
In the classical limit, $Z$ is equal to one and the function $f(Z)$ can be parametrized by $f(Z)$ = 
$\left( 1- Z\right)^\rho$, where $\rho$ is a positive parameter. The parameter $\rho$ is probably equal to 1. 
If $\rho$ = 1, we recover the superpotential for $N_f$ = $N_c$+1 given by
$W_{\rm eff}$ = $S
\left\{
\ln
\left[
\left( 
\det(T)\right.\right.\right.$-$\left.\left.BT{\bar B}
\right)
/S\Lambda^{3N_c-N_f} 
\right]$
+ $\left.1
\right\}$\cite{Seiberg}.
 
\section{Strategy}

To examine dynamical properties of our superpotential, we
\begin{enumerate}
\item first go to slightly broken SUSY vacuum, where 
symmetry behavior of the superpotential is more visible.
\item use universal scalar masses of $\mu_L$ and $\mu_R$, which respect global symmetry and  only break SUSY.
\item check the consistency with SQCD in its SUSY limit, after determining the SUSY-broken vacuum.
\end{enumerate}
\noindent
The SQCD defined in the SUSY limit of the so-obtained SUSY-broken SQCD should exhibit the consistent 
anomaly-matching property and yield the compatible result with instanton calculus.

Let $\pi_i$ be $\langle 0\vert T^i_i \vert 0\rangle$, $\pi_\lambda$ be $\langle 0\vert S\vert 0\rangle$  
 and $z$ be $\langle 0\vert Z \vert 0\rangle$. 
Since the dynamics requires that some of the $\pi$ acquire non-vanishing VEV's, 
suppose that one of the $\pi_i$ ($i$=1 $\sim$ $N_f$) develops a VEV, and let this be  
labeled by $i$ = $1$: $|\pi_1|$ $\sim$ $\Lambda^2$.  This VEV is determined by 
solving  $\partial V_{\rm eff}/\partial\pi_i$ = 0, yielding
\begin{eqnarray}\label{Eq:Tab}
&&G_TW_{{\rm eff};a}^\ast
\frac{\pi_\lambda}{\pi_a} \left( 1-\alpha \right) 
 =  G_SW_{{\rm eff};\lambda}^\ast \left( 1-\alpha \right)
\nonumber \\
&&+\beta X +M^2\big| \frac{\pi_a}{\Lambda}\big|^2, ~~~~~~\left(a = 1{\sim}N_c\right), 
\end{eqnarray}
where $\alpha$ = $zf^\prime (z)/f(z)$; 
$\beta$ = $z\alpha^\prime$; $M^2$ = $\mu_L^2$ + $\mu_R^2$ + $G_T^\prime\Lambda^2\sum_{i=1}^{N_f} \big| 
W_{{\rm eff};i}\big|^2$; $X$ = $G_T\sum_{a=1}^{N_c}W_{{\rm eff};a}^\ast\left(\pi_\lambda /\pi_a\right) - 
G_B\sum_{x=B,{\bar B}}W_{{\rm eff};x}^\ast$ $\left(\pi_\lambda / \pi_x\right)$; 
$W_{\rm eff;i(\lambda)}$ = $\partial W_{\rm eff}/\partial\pi_{i(\lambda)}$; $G$'s come from 
field-dependent K${\ddot {\rm a}}$hler potentials. 
The SUSY breaking effect is specified by $(\mu_L^2+\mu_R^2)\vert \pi_1 \vert^2$ through $M^2$   
because of $\pi_1$ $\neq$ 0.

Without knowing the details of solutions to these equations, we can find that  
\begin{eqnarray}
&&\bigg| \frac{\pi_a}{\pi_1} \bigg|^2 = 1 + 
\frac{\frac{M^2}{\Lambda^2}(\big| \pi_1\big|^2-\big| \pi_a\big|^2)} {G_SW_{{\rm eff};\lambda}^\ast \left( 1-\alpha \right) +
\frac{M^2}{\Lambda^2}\big| \pi_a\big|^2+\beta X},
\end{eqnarray}
which cannot be satisfied by $\pi_{a\neq 1}$ = 0.  In fact, $\pi_{a\neq 1}$ = $\pi_1$ is a solution to 
this problem, leading to $|\pi_a|$ = $|\pi_1|$.  Then, you can see the emergence of 
the vectorial $SU(N_c)$ symmetry. 

\section{Symmetry Breaking}

Using the input of $\vert \pi_{i=1\sim N_c}\vert$ $\equiv$ $\Lambda^2_T$ $\sim$ $\Lambda^2$ just obtained, 
we reach the solutions given by 
$\vert \pi_B\vert$ = $\vert \pi_{\bar B}\vert$ $\equiv$ $\Lambda^{N_c}_B$ $\sim$ $\Lambda^{N_c}$, 
$\vert \pi_{i=N_c+1\sim N_f}\vert$ = $\epsilon\vert \pi_{1\sim N_c}\vert$ and 
$\vert \pi_\lambda\vert$ $\sim$ $\epsilon^{1+\frac{\rho}{N_f-N_c}}\Lambda^3$. 
Notice that $\pi_i$ ($i$ = $N_c$+1 $\sim$ $N_f$) and $\pi_\lambda$ accompany the factor $\epsilon$.  
This parameter $\epsilon$, defined to be $\vert 1-z\vert$, measures the SUSY breaking effect.
So, taking the SUSY limit with $\epsilon$ $\rightarrow$ 0, we reach the SUSY vacuum specified by these VEV's.
The solutions clearly show the presence of vectorial $SU(N_c)$ symmetry and chiral $SU(N_f-N_c)$ symmetry.  
The resulting breaking pattern is described by 
$G$ = $SU(N_f)_L$ $\times$ $SU(N_f)_R$ $\times$ $U(1)_V$ $\times$ $U(1)_A$ down to 
$H$ = $SU(N_c)_{L+R}$ $\times$ $SU(N_f-N_c)_L$ $\times$ $SU(N_f-N_c)_R$ $\times$ $U(1)^\prime_V$ 
$\times$ $U(1)^\prime_A$.

We find consistent anomaly-matching property due to the emergence of the Nambu-Goldstone superfields 
associated with $G$ $\rightarrow$ $H$, where massless bosons responsible for the anomalies of the broken 
part, $G/H$, and massless fermions for those of the unbroken part, $H$.  
Therefore, the anomaly-matching is a purely dynamical consequence.  
We have further checked that our superpotential is  consistent with holomorphic decoupling and 
instanton calculus for SQCD with $N_f$ = $N_c$ reproduced by massive quarks with flavors of $SU(N_f-N_c)$.  
The detailed description can be found in the literature\cite{Yasue}.
 
\section{Summary}

Dynamical breakdown of chiral symmetries are shown to be determined by the effective superpotential: 
$W_{\rm eff}$ = 
$S$ $\{ \ln [$ $S^{N_c-N_f}$ ${\rm det}(T)$ $f(Z)$ $\Lambda^{N_f-3N_c}]$ + $N_f-N_c\}$ 
with $f(Z)$ dynamically determined to be $\left( 1-Z\right)^\rho$ ($\rho$ $>$ 0),
It will be realized at least in SQCD with $N_c$+2 $\leq$ $N_f$ $\leq$ 3$N_c$/2. 
This superpotential exhibits 
\begin{enumerate}
\item holomorphic decoupling property,
\item spontaneously breakdown of chiral $SU(N_c)$ symmetry and restoration of chiral $SU(N_f-N_c)$ 
symmetry described by $SU(N_f)_L\times SU(N_f)_R\times  U(1)_V \times U(1)_A 
\rightarrow SU(N_c)_{L+R} \times SU(N_f-N_c)_L \times SU(N_f-N_c)_R \times U(1)^\prime_V 
\times U(1)^\prime_A$,
\item consistent anomaly-matching property due to the emergence of the Nambu-Goldstone superfields, and 
\item correct vacuum structure for $N_f$ = $N_c$ reproduced by instanton contributions when all 
quarks with flavors of $SU(N_f-N_c)$ become massive. 
\end{enumerate}

In this end, we have two phases in SQCD: 
one with chiral $SU(N_f)$ symmetry for $\ddash$magnetic" quarks and the other 
with spontaneously broken chiral $SU(N_f)$ symmetry for the Nambu-Goldstone superfields. 
This situation can be compared with the case in the ordinary QCD with two flavors: one with proton and neutron and the other with pions.

Finally, I mention related three works here, which are characterized by
\begin{itemize}
\item Dynamical evaluations of condensates\cite{GapEquation}, 
\item Instable SUSY vacuum in the "magnetic" phase\cite{Instability},
\item Slightly different effective superpotential in the "electric" phase\cite{S}.
\end{itemize}
\noindent
All these works indicate spontaneous chiral symmetry breaking in the $\ddash$electric" phase.

\end{multicols}


\begin{thebibliography}{99}
%
\bibitem{Seiberg} N. Seiberg, Phys. Rev. D {\bf 49}, 6857 (1994); 
Nucl. Phys. B {\bf 435}, 129 (1995).
%
\bibitem{CoulombPhase} K. Intriligator and N. Seiberg, Nucl. Phys. {\bf B431} (1994) 551;
P.C. Argyres, M.R. Plesser and A.D. Shapere, Phys. Rev. Lett. {\bf 75} (1995) 1699;
A. Hanany and Y. Oz, Nucl. Phys. {\bf B452} (1995) 283.
%
\bibitem{BrokenN_2} R.G. Leigh and M.J. Strassler, Nucl. Phys. {\bf B447} (1995) 95;
P.C. Argyres, M.R. Plesser and N. Seiberg, Nucl. Phys. {\bf B471} (1996) 159;
M.J. Strassler, Prog. Theor. Phys. Suppl. {\bf No.123} (1996) 373;
N.Evans, S.D.H. Hsu, M. Schwetz and S.B. Selipsky, Nucl. Phys. Proc. Suppl {\bf 52A} (1997) 223;
P.C. Argyres, Nucl. Phys. Proc. Suppl {\bf 61A} (1998) 149;
T. Hirayama, N. Maekawa and S. Sugimoto, Prog. Theor. Phys. {\bf 99} (1998) 843.
%
\bibitem{Yasue} Y. Honda and  M. Yasu${\grave {\rm e}}$, Prog Theor. Phys. {\bf 101}, 971 (1999); 
Phys. Lett. B {\bf 466}, 244 (1999).
%
\bibitem{tHooft} G. 't Hooft, in {\em Recent Development in Gauge Theories}, 
Proceedings of the Cargese Summer Institute, Cargese, France, 1979, edited 
by G. 't Hooft  {\em et al.}, 
NATO Advanced Study Institute Series B: Physics Vol. 59 (Plenum Press, New York, 1980).
%
\bibitem{AnomalyMatch} 
T. Banks, I. Frishman, A. Shwimmer and S. Yankielowicz, Nucl. Phys. {\bf B177}, 157 (1981). 
%
\bibitem{VY} G. Veneziano and S. Yankielowicz, Phys. Lett.  {\bf 113B}, 321 (1983); 
T. Taylor, G. Veneziano and S. Yankielowicz, Nucl. Phys. {\bf B218}, 493 (1983). 
%
\bibitem{Weff} 
A. Masiero, R. Pettorino, M. Roncadelli and G. Veneziano, Nucl. Phys. 
{\bf B261} (1985) 633; M. Yasu${\grave {\rm e}}$, Phys. Rev. D {\bf 35}  (1987) 355 and D 
{\bf 36} (1987) 932; Prog Theor. Phys. {\bf 78} (1987) 1437.
%
\bibitem{GapEquation} 
T. Appelquist, A. Nyffeler and S.B. Selipsky, Phys. Lett. B {\bf 425} (1998) 300.
%
\bibitem{Instability}  N. Arkani-Hamed and R. Rattazzi,  Phys. Lett. B {\bf 454} (1999) 290. 
%
\bibitem{S}  P.I. Pronin and K.V. Stepanyantz, hep-th/9902163 (Feb., 1999).
%
\end{thebibliography}
\end{document}